\documentclass[letterpaper,11pt]{article}
\pdfoutput=1 
\usepackage{xcolor}
\usepackage{jheppub}
\usepackage{mathtools}
\usepackage{tikz}
\usepackage{braket}
\usepackage{appendix}
\usepackage{tabularx}

\usepackage{amsmath}
\usepackage{simplewick}

\renewcommand{\b}[1]{\textbf{#1}}
\usetikzlibrary{tikzmark}
\newcommand*\circled[1]{\tikz[baseline=(char.base)]{
    \node[shape=circle,draw,inner sep=2pt] (char) {#1};}}
\newcommand{\al}{\alpha}
\newcommand{\an}[1]{\left\langle#1\right\rangle}
\newcommand\D{\Delta}
\newcommand{\dhat}{\Hat{\mathcal{D}}}
\newcommand\e{\varepsilon}
\newcommand{\om}{\omega}
\newcommand{\mA}{\mathcal{A}}
\newcommand{\mAt}{\tilde{\mathcal{A}}}
\newcommand\mO{\mathcal{O}}

\newcommand{\sq}[1]{\left[#1\right]}

\newcommand{\zb}{\bar{z}}
\newcommand{\nn}{\nonumber}

\newcommand{\veb}{\bar{\varepsilon}}
\newcommand{\hb}{\bar{h}}

\title{Celestial soft currents at one-loop and their OPEs}

\author[a]{Rishabh Bhardwaj}
\author[b]{and Akshay Yelleshpur Srikant}

\affiliation[a]{Department of Physics,
	Brown University,
	Providence, RI 02912, USA}
\affiliation[b]{Mathematical Institute, University of Oxford,
Andrew Wiles Building, Radcliffe Observatory Quarter,
Woodstock Road, Oxford, OX2 6GG, UK}
\emailAdd{rishabh\_bhardwaj@brown.edu}\emailAdd{Akshay.YelleshpurSrikant@maths.ox.ac.uk}

\abstract{Conformally soft operators and their associated soft theorems on the celestial sphere encode the low energy behaviour of bulk scattering amplitudes. They lead to an infinite dimensional symmetry algebra of the celestial CFT at tree-level. In this paper, we introduce new operators in the celestial CFT in order to extend the definition of conformally soft currents to include one-loop effects. We then compute their OPEs with other operators in the theory. We also examine new subtleties that arise in defining OPEs of two conformally soft operators. We elucidate the connection between the new operators and loop corrected soft theorems in the bulk. Finally, we conclude by demonstrating how these operators fit into the framework of a logarithmic CFT.}

\begin{document} 
\maketitle
\flushbottom

\section{Introduction}
The Celestial Holography program \cite{Pasterski:2017ylz, Pasterski:2017kqt} aims to reformulate scattering amplitudes in asymptotically flat spacetimes in terms of correlators of a 2D conformal field theory living on the Celestial Sphere at null infinity. This theory - dubbed Celestial Conformal Field Theory (CCFT) - is essentially defined in terms of the bulk S-Matrix elements and currently lacks an independent definition \footnote{Some notable exceptions are toy models of self-dual gravity in \cite{Costello:2022upu, Bu:2022iak}}. Progress towards achieving this requires a knowledge of its spectrum and the associated OPE coefficients. Symmetries, in addition to the eponymous conformal symmetries, when used appropriately, typically enhance our computational prowess. This has been the central motivation of many papers devoted to uncovering symmetries of CCFT and to extracting the spectrum and OPE data directly from scattering amplitudes. The collinear limit of scattering amplitudes in the bulk probes the OPE limit in the boundary CCFT \cite{Fan:2019emx, Pate:2019lpp, Himwich:2021dau}. The leading terms in the collinear limit of amplitudes are governed by the three particle interactions and are universal meaning that they are independent of the other particles involved in the scattering process \cite{Mangano:1990by, Berends:1987me}. This information can be used to compute some OPE coefficients which contain non-trivial information about the spectrum and symmetries of CCFT. Further information about the spectrum can be gained from conformal block expansions of CCFT four point functions \cite{Fan:2021isc, Atanasov:2021cje, Nandan:2019jas, De:2022gjn, Hu:2022syq}.\\

Singularities of the OPE coefficients are predictive of universal singularities of CCFT correlators which dovetails with expectations from soft theorems in gauge theory and gravity \cite{Weinberg:1965nx, Cachazo:2008hp, Casali:2014xpa}. Together, they imply that the CCFT spectrum contains an infinite tower of operators - called conformally soft operators - whose dimensions are integers less than 2 \cite{Pate:2019mfs, Puhm:2019zbl, Donnay:2018neh}. Such operators have been shown to form infinite dimensional symmetry algebras - called the $\mathcal{S}$-algebra in Yang-Mills \cite{Guevara:2021abz} and the wedge subalgebra of $Lw_{1+\infty}$-algebra in gravity \cite{Strominger:2021lvk}. Most of these results, particularly in the case of Yang-Mills theory, are restricted to tree-level, meaning that they are derived from properties of tree-level amplitudes. Both the collinear limit \cite{Kosower:1999xi, Bern:1999ry, Kosower:1999rx, Bern:1995ix, Bern:1994zx} and the soft limit \cite{Bern:2014oka, Bern:1998sc, Bianchi:2014gla, He:2014bga, Elvang:2016qvq, Sahoo:2018lxl, Krishna:2023fxg} of amplitude receive one-loop corrections in Yang-Mills theory implying that the symmetry algebra found at tree-level is modified in some way. \footnote{In gravitational theories, such effects are intimately tied to the fundamental properties of CCFT via the stress tensor. This has been explored in \cite{He:2017fsb, Pasterski:2022djr, Donnay:2022hkf}.} Moreover, since our approach to Celestial Holography is a bottom up one, it is crucial that these effects can be consistently accommodated into a CFT-like framework. This informs the twofold motivations for this paper. The first of these is to build on the results of \cite{Bhardwaj:2022anh} and continue laying the groundwork necessary for understanding the effect of loop corrections on the infinite dimensional symmetry algebras in Yang-Mills theory. Secondly, it is to understand how the spectrum of CCFT must be modified to accommodate the loop corrections in the bulk. \\

The paper is structured as follows. We start Section [\ref{sec:review}] by reviewing one-loop corrections to the collinear limit and the corresponding OPE coefficients. We then present new formulas for OPE coefficients which include all right moving descendants. These formulas are implied by conformal symmetry from the ones derived in \cite{Bhardwaj:2022anh}. In Section [\ref{sec:confsoftdef}], we explain how to define conformally soft operators in the presence of loop corrections and explore the OPEs of these operators with other hard operators in Section[\ref{sec:hardsoftOPE}]. We move to discussing OPEs of two conformally soft operators. Here, we find that they are ill defined at loop level due to ambiguities involving orders of limits. We demonstrate this explicitly in Section [\ref{sec:softsoftOPE}]. In Section [\ref{sec:OPEsofttheorem}], we explain how the loop corrected soft theorems can be phrased in terms of the conformally soft operators defined in the previous sections and in Section [\ref{sec:logcft}], we explore the log-CFT like structure of conformally soft operators and end with some discussion in Section [\ref{sec:discussion}].\\

\emph{Note: While this work was in progress, we became aware of \cite{Krishna:2023ukw} which has some overlapping results, in particular the formula \ref{eq:++OPEblockfinal} for the OPE including descendants and parts of Section [\ref{sec:confsoftdef}]. }


\section{One-loop OPEs in Yang-Mills theory}
\label{sec:review}
The focus of this paper is on celestial amplitudes of gluons at one-loop. In the interest of increasing readability of expressions, we will work in the context of supersymmetric Yang-Mills theory. In addition to simplifying many of the formulas that appear in this paper, it is also of interest in its own right \cite{Ball:2023qim}. Most of the results of this paper generalize in a straightforward manner to non supersymmetric Yang-Mills and we will comment on this when relevant. Celestial amplitudes (denoted by $\tilde{A}_n$) are Mellin transforms of momentum space scattering amplitudes (denoted by $A_n$) and thus inherit many of their structures. There are two structures which are of primary relevance to this paper. The first of these is the perturbative expansion in the coupling constant
\begin{align}
    \label{eq:loopexpansion}
   A_n \left(p_1,\sigma_1; \dots p_n,\sigma_n\right) = g^{n-2} \sum_{L=0}^{\infty} \al^L A_n^{(L)}\left(p_1,\sigma_1; \dots p_n,\sigma_n\right),
\end{align}
where $g$ is the Yang-Mills coupling constant, $\al = a \left(4\pi e^{-\gamma_E}\right)^{\e}$ with $a= \frac{g^2 N_c}{8\pi^2}$ the 't Hooft coupling, $\epsilon = 4-D$ the dimensional regularization parameter, $\left\lbrace p_1, \dots p_n \right\rbrace$ the momenta of the external gluons, $\sigma_i = \pm 1$ their helicities. The second is the decomposition of these amplitudes into colour ordered partial amplitudes. At tree-level, this takes the form
\begin{align}
    \label{eq:tracedecomptree}
    &A_n^{(0)} = g^{n-2} \sum_{s \in S_n/Z_n}\text{Tr} \left[T^{a_{s\left(1\right)}} \dots T^{a_{s\left(n\right)}}\right] \mA_n^{(0),a_{s(1)} \dots a_{s(n)}},
\end{align}
where $T^{a_i}$ are the generators of $SU(N_c)$ in the fundamental representation and $S_n/Z_n$ are the non cyclic permutations of $\left\lbrace1, \dots, n\right\rbrace$. We've suppressed the kinematic dependence on both sides. $ \mA_n^{(0),a_1 \dots a_n}$ are gauge invariant, colour ordered partial amplitude given by the Parke-Taylor formula. This decomposition is modified for one-loop amplitudes and reads
\begin{align}
    \label{eq:tracedecomploop}
        &A_n^{(1)} = g^{n-2}\al\sum_{\mathcal{R}} n_\mathcal{R}\sum_{c=1}^{\left\lfloor \frac{n}{2}\right\rfloor+1}\sum_{s \in S_n/S_{n;c}}\text{Gr}_{n;c}\left(s\right)\mA_{n;c}^{(1),a_{s(1)} \dots a_{s(n)}},
\end{align}
where $n_{\mathcal{R}} = 1$ for particles in the adjoint representation and $n_{\mathcal{R}} =\frac{1}{N_c}$ for particles in the fundamental representation. Furthermore,
\begin{align}
    \text{Gr}_{n;c} \left(s\right) = \begin{cases} \text{Tr} \left[T^{a_{s\left(1\right)}} \dots T^{a_{s\left(n\right)}}\right] \qquad &c = 1 \\
    \frac{1}{N_c}\text{Tr} \left[T^{a_{s\left(1\right)}} \dots   T^{a_{s\left(c-1\right)}}\right]\text{Tr}\left[T^{a_{s\left(c\right)}} \dots T^{a_{s\left(n\right)}}\right] \qquad &c>1,
    \end{cases}
\end{align}
and $S_{n;c}$ is the set of permutations which leave $\text{Gr}_{n;c}$ invariant. It can be shown that $\mA_{n;c}$ for $c>1$ can be expressed as linear combinations of the various $\mA_{n;1}$. Thus, is suffices to focus only on this trace structure. We will often omit the colour indices entirely but it should be understood that we are always working with the colour ordered amplitude. Combining the perturbative expansion in (\ref{eq:loopexpansion}) and the colour decomposition allows us to define colour ordered Celestial amplitudes
\begin{equation}
\label{eq:celampdef}
    \tilde{\mathcal{A}}_n^{(L)} \left(\D_1,\sigma_1;\dots; \D_n,\sigma_n\right) =\int_0^{\infty}\prod_{i=1}^n\,d\omega_i\,\, \omega_i^{\Delta_{i}-1}\,\mathcal{A}_n^{(L)}\left(p_1,\sigma_1; \dots; p_n,\sigma_n\right),
\end{equation}
where we have parameterized an outgoing massless four-momentum of an external gluon as
\begin{equation}
    p_i^{\mu}= \frac{\Lambda \, \omega}{2}\left(1+z_i \zb_i,z_i+\bar{z}_i,-i(z_i-\bar{z}_i),1-z_i\bar{z}_i\right)\label{eq:momentaparameterization}.
\end{equation}
Here $\Lambda$ is a new parameter that has dimensions of energy and $\omega$ is kept dimensionless for the Mellin integral to make sense dimensionally. The expression as written, is valid in a $(+,-,-,-)$ signature bulk spacetime and expressions for other signatures can be obtained by suitable analytic continuation. Throughout, we will assume all particles to be outgoing. The following spinor helicity variables and their Lorentz invariant contractions will be of use later in the paper:
\begin{align}
\label{eq:anglesquare}
    &\lambda_i = \sqrt{\Lambda\omega_i}\begin{pmatrix} 1 \\ z_i\end{pmatrix}, \qquad \tilde{\lambda}_i = \sqrt{\Lambda\omega_i}\begin{pmatrix} 1 \\ \zb_i\end{pmatrix}, \\
    &\an{ij} = \Lambda\sqrt{\omega_i \, \omega_j} z_{ij}\,, \qquad \sq{ij} = \Lambda \sqrt{\omega_i \, \omega_j}\zb_{ij}\,,\nonumber
\end{align}
where $z_{ij} = z_i - z_j$. We will identify the celestial amplitude with a CCFT correlator
\begin{align}
    \label{eq:CCFT}
    \tilde{\mathcal{A}}_n  \left(\D_1,\sigma_1;\dots; \D_n,\sigma_n\right) = \an{\mO_{\D_1,\sigma_1}\left(z_1,\zb_1\right) \dots \mO_{\D_n,\sigma_n}\left(z_n,\zb_n\right)}.
\end{align}
This identification implies that the OPEs can be computed from collinear limits. We are interested in the OPEs of gluon operators in supersymmetric Yang-Mills theories. These OPEs are special cases of the more general ones studied in \cite{Bhardwaj:2022anh}. In the collinear limit, $p_1 \cdot p_2 \to 0$, tree-level amplitude behaves as
\begin{equation}
    \label{eq:collineartree}
    \mA_n^{(0)} \left(p_1,+;p_2,+; \dots \right) \xrightarrow[]{1\parallel2}\, \frac{1}{\sqrt{t(1-t)}\braket{12}}\,\mA_{n-1}^{(0)}(P,+;\dots)\,.
\end{equation} 
where $p_1 = t P, p_2 = (1-t)P$ is a parametrization of the collinear limit. The above behaviour is universal in the sense that it is independent of the identities of particles $3, \dots, n$. One-loop gluon amplitudes also exhibit universal behaviour ~\cite{Bern:1994zx, Bern:1996je, Bern:1995ix, Mangano:1990by, Dixon:1996wi, Bern:1999ry, Kosower:1999rx, Kosower:1999xi, Bern:2004cz}. This is particularly simple in supersymmetric theories where it takes on the form \footnote{This can be obtained from the equations in \cite{Bhardwaj:2022anh} by restricting to the case of equal number of bosons and fermions.}
\begin{multline}
\label{eq:collinearloop}
    \mA_n^{(1)} \left(p_1,+;p_2,+; \dots \right)  \overset{1\parallel 2}{\longrightarrow}\, \frac{1}{\sqrt{t(1-t)}\braket{12}}\Big[\,\mathcal{A}_{n-1}^{(1)} \left(P,+; \dots\right) +G^n\,\mathcal{A}_{n-1}^{(0)} \left(P,+; \dots\right)\Big],
\end{multline}
with
\begin{align}
\label{eq:nonfact}
    G^n&=\hat{c}_{\Gamma}\Bigg[-\frac{1}{\epsilon^2}\Bigg(\frac{\mu^2}{-s_{12}t(1-t)}\Bigg)^{\epsilon}+2\ln(t)\ln(1-t)-\frac{\pi^2}{6}\Bigg]+\mathcal{O}(\epsilon)\,,
\end{align}
where $s_{12} = \an{12}\sq{21}$ and
\begin{equation}
\label{eq:chat}
    \hat{c}_{\Gamma} = \frac{e^{\epsilon\gamma}}{2}\frac{\Gamma(1+\epsilon)\Gamma^2(1-\epsilon)}{\Gamma(1-2\epsilon)}\,,
\end{equation}
and $\epsilon = \frac{4-D}{2}$ is the dimensional regularization parameter. The OPE follows from Mellin transforming the above relation. At this stage, it is useful to deviate from the analysis of \cite{Bhardwaj:2022anh} by considering the holomorphic collinear limit. This is possible by either formally complexifying the celestial sphere variables or by working in split signature spacetime. In either case, the holomorphic collinear limit refers to $z_{12} \to 0$ and $\zb_{12}$ arbitrary. The details of this derivation are very similar to \cite{Bhardwaj:2022anh, Guevara:2021abz} and we merely present the result \footnote{We have dropped all terms corresponding to UV renormalization everywhere in this paper since they do not involve independent coefficients. It is straightforward to reintroduce them at any stage.}
\begin{equation}
\label{eq:loopOPEblock}
      \mO^a_{\D_1,+} \, \mO^b_{\D_2,+} \sim \frac{g^2 i f^{abc}}{z_{12}} \left[\left(1-\alpha\frac{\pi^2}{6}\right) X^c_1 -\frac{\alpha\hat{c}_{\Gamma}}{\epsilon^2 }\left(\frac{-1}{\left|z_{12}\right|^2}\right)^{\epsilon} X^c_2 +2\alpha\hat{c}_{\Gamma} X^c_3\right],
\end{equation}
where $f^{abc}$ is the $SU(N)$ structure constant, $\left|z_{12} \right|^2 = z_{12}\zb_{12}$,
\begin{align}    
\label{eq:Cdefs}
    X^c_1 = \int_0^1 d\mathcal{T}\left(\D_1,\D_2\right)&\mO^c_{\D} \left(z_2, \zb_2+t \zb_{12} \right), \quad X^c_3 = \partial_{\D_1} \partial_{\D_2}\int_0^1 d\mathcal{T}\left(\D_1,\D_2\right) \mO^c_{\D} \left(z_2, \zb_2+t \zb_{12} \right),\nonumber\\
    &X^c_2 =\int_0^1 d\mathcal{T}\left(\D_1-2\epsilon,\D_2-2\epsilon\right)\mO^c_{\D-2\epsilon} \left(z_2, \zb_2+t \zb_{12} \right),
\end{align}
and we have defined a new measure $d\mathcal{T}\left(a, b\right)$ and conformal dimension $\D$ for brevity. These are
\begin{equation}
    \label{eq:dtdeltadef}
    d\mathcal{T}\left(a, b\right) = dt \,t^{a-2} (1-t)^{b-2} \qquad  \D = \D_1+\D_2-1.
\end{equation}
In writing (\ref{eq:loopOPEblock}), we have suppressed the coordinate dependence on both sides and set $\Lambda = \mu$. It should be understood that the operators are inserted at $z_1, z_2$ and the coordinate dependence of the RHS of the OPE is explicit in (\ref{eq:Cdefs}). (\ref{eq:loopOPEblock}) constitutes an OPE block in the $\zb$ coordinate, i.e. it includes all the $\zb$ descendants. This can be easily seen by Taylor expanding each of the integrands in (\ref{eq:Cdefs}) in $t$ and performing the resulting integrals:
\begin{align}
     &X^c_1 = \sum_{m=0}^{\infty} \frac{\zb_{12}^m}{m!} B\left(\D_1-1+m,\D_2-1\right) \bar{\partial}^m\mO^c_{\D} \left(z_2, \zb_2\right),\nonumber\\
    &X^c_2 = \sum_{m=0}^{\infty} \frac{\zb_{12}^m}{m!} B\left(\D_1-1+m-2\epsilon,\D_2-1-2\epsilon\right) \bar{\partial}^m\mO^c_{\D-2\epsilon} \left(z_2, \zb_2\right),\\
    &X^c_3 = \sum_{m=0}^{\infty} \frac{\zb_{12}^m}{m!}  \partial_{\D_1} \partial_{\D_2}B\left(\D_1-1+m,\D_2-1\right) \bar{\partial}^m\mO^c_{\D} \left(z_2, \zb_2\right),\nonumber
\end{align}
where $\bar{\partial} \equiv \frac{\partial}{\partial \zb_2}$. This OPE is invariant under conformal transformations of the $\zb$ coordinate as shown in Appendix [\ref{app:Confinvarcheck}]. We can now connect with the results of \cite{Bhardwaj:2022anh}, by performing a series expansion in $\epsilon$ and keeping terms upto $\mO(1)$. This gives
\begin{align}
\label{eq:++OPEblockfinal}
    \mO^a_{\D_1,+} \mO^b_{\D_2,+} &\sim \frac{ig^2f^{abc}}{z_{12}} \Big[1+\frac{a}{2}\Big( C^{(1)}_{0,+} + C^{(1)}_{1,+} \mathcal{\hat{D}}_{12} + C^{(1)}_{2,+}\mathcal{\hat{D}}_{12}^2\Big)\Big]X_1^c  \\
 & = \frac{ig^2f^{abc}}{z_{12}} \Big[1+\frac{a}{2}\Big( C^{(1)}_{0,+} + C^{(1)}_{1,+} \mathcal{\hat{D}}_{12} + C^{(1)}_{2,+}\mathcal{\hat{D}}_{12}^2\Big)\Big] \sum_{m=0}^{\infty} C_{+}^{(0)}\mO^c_{\D+m,+}\left(z_2,\zb_2\right). \nonumber
\end{align}
This formula has also been derived in \cite{Krishna:2023ukw}. We have suppressed all arguments of the OPE coefficients - a practice we will maintain throughout this paper. In particular, note that $C_+^{(0)}$ is a function of $m$ and must be retained within the summation. Explicit expressions for these are
\begin{align}
    &C^{(1)}_{0,+} = H_{0,+}^{(1)} + S_{0,+}^{(1)}\,, \qquad S_{0,+}^{(1)} =-\frac{1}{\epsilon^2}+\frac{c_E}{\epsilon}-\frac{c_E^2}{2}, \qquad  H_{0,+}^{(1)} = -\frac{\pi^2}{12}+2\partial_{\D_1}\partial_{\D_2}, \label{eq:++OPEfinalcoeffs}\\
& C^{(1)}_{1,+} = 2\left(\frac{1}{\epsilon}-c_E\right), \qquad C^{(1)}_{2,+} = -2,  \qquad C_{+}^{(0)} = \text{B}\left(\D_1-1+m,\D_2-1\right),\nonumber
\end{align}
where $\mathcal{\hat{D}}_{ij}$ is the ``covariant derivative'' operator
\begin{equation}
\label{eq:covariantderivative}
    \mathcal{\hat{D}}_{ij}=\partial_{\Delta_i}+\partial_{\Delta_j}+\partial_{\Delta}+\frac{1}{2}\ln\left(-\left|z_{ij}\right|^2\right).
\end{equation}
This is a natural object that transforms covariantly under conformal transformation in the OPE limit $z_{ij},\zb_{ij}\rightarrow 0$. We must append the definition (\ref{eq:covariantderivative}) with following prescription: prior to differentiation, assume that $\D$ is independent of $\D_1, \D_2$ and set $\Delta = \Delta_1 + \Delta_2 - 1$ only thereafter. This implies that $\partial_{\D_i}, \partial_{\D_j}$ do not act on $\mO_{\D}$. The following example is illustrative of this rule :
\begin{align}
     \nonumber\mathcal{\hat{D}}_{12} \left(C^{(0)}_{+} \left(m \right) \mO^c_{\D,+}\left(z_2, \zb_2 \right)\right) 
     =&  \,\mO^c_{\D,+}\left(z_2, \zb_2 \right)\left(\partial_{\Delta_1}+\partial_{\Delta_2} + \frac{1}{2}\ln\left(- z_{ij}\zb_{ij}\frac{\Lambda^2}{\mu^2}\right)\right)C^{(0)}_{+} \\
     &+C^{(0)}_{+} \partial_{\Delta}\mO^c_{\D,+}\left(z_2, \zb_2 \right)\Big|_{\D = \D_1+\D_2-1}.
\end{align}
The formula for the OPE in (\ref{eq:++OPEblockfinal}) is the analog of the formula presented in \cite{Bhardwaj:2022anh} but enhanced to include all $\zb$ descendants. More concretely, we reproduce the supersymmetric version of (3.35) of \cite{Bhardwaj:2022anh} if we keep only the $m=0$ term in (\ref{eq:++OPEblockfinal}).\\

Loop integrals in Yang-Mills theory are IR divergent and need to be regularized. Dimensional regularization has been employed in writing all the expressions presented thus far in this paper. This divergence carries over to the collinear limits and consequently the OPEs. The coefficients $S^{(1)}_{0,+}, C_{1,+}^{(1)}$ in (\ref{eq:++OPEfinalcoeffs}) diverge as $\epsilon \to 0$. We can extract an IR-finite OPE from \ref{eq:++OPEblockfinal}) by taking inspiration from the hard-soft factorization of gauge theory amplitudes and introducing the following decomposition of the gluon operators \cite{Magnea:2021fvy}:
\begin{align}
      \mathcal{O}^{a}_{\Delta,+}(z,\bar{z})=[\b{V}_{\kappa}(z,\bar{z})]^{a}_{a'}H^{a'}_{\Delta',+}(z,\bar{z})\,.\label{eq:decompoperator}
\end{align}
At this stage, it is worth pausing here to sort out some terminology while simultaneously highlighting the salient features of the decomposition above. The operators $[\b{V}_{\kappa}(z,\bar{z})]^{a}_{a'}$ reproduce the IR divergent part of the OPE \footnote{Here too, we have dropped all terms corresponding to UV renormalization effects.}
\begin{equation}
\label{eq:vertexOPE2}
      f^{a'b'c'}[\b{V}_{\kappa_1}]^{a}_{a'}\,\, [\b{V}_{\kappa_2}]^{b}_{b'} \sim  f^{abc} \left(1+\frac{a}{2}S_{0,+}^{(1)}+\frac{a}{4}C_{1,+}^{(1)}\log \left(-\left|z_{12}\right|^2\right) \right)[\b{V}_{\kappa}]_{c'}^{c}\, ,
\end{equation}
with $S_{0,+}^{(1)}, C_{1,+}^{(1)}$ defined in (\ref{eq:++OPEfinalcoeffs}). This renders the OPE of the $H^{a'}_{\Delta',\sigma}(z,\bar{z})$ operators finite. In \cite{Bhardwaj:2022anh}, these were termed ``hard'' operators which was meant to indicate that they do not include the IR divergences arising from soft and collinear configurations of \emph{loop} momenta. In the context of this paper, this terminology is potentially confusing since we will consider taking the momenta of the \emph{external} particles soft (and the corresponding conformally soft limits). The OPEs involving $H_{\D',+}^a$ are certainly singular in this limit, as they are manifestly so in the similar collinear limit. We will thus refer to the $H^{a'}_{\Delta',+}$ as IR finite operators. Finally, note that the conformal dimension $\D'$ of $H^{a'}_{\Delta',+}(z,\bar{z})$ is shifted from that of $\mO_{\D,+}$  and the two are related by  $\D' = \D + \frac{a}{2}C_{1,+}^{(1)}$. Since the majority of this paper will deal with OPEs of the IR finite operators, we will disregard this distinction and drop the prime over the dimension of the IR finite operators. Their OPE is,
\begin{multline}
\label{eq:hardOPEfinal}
    H^{a}_{\Delta_1,+} \, H^{b}_{\Delta_2,+}
     \sim\frac{ig^2f^{abc}}{z_{12}}\left[1
      -\frac{a \,\pi^2}{24}+a\,\partial_{\D_1}\partial_{\D_2}-a\,\dhat_{12}^2\right]\int_0^1 d\mathcal{T}\left(\D_1,\D_2\right) H_{\D,+}^{c}\left(z_2,\zb_2+t \zb_{12}\right) 
\end{multline} 
We can now explicitly evaluate all of these integrals by Taylor expanding the operator in the integrand in $t$. This gives
\begin{align}
\label{eq:IRfiniteOPE}
    H^{a}_{\Delta_1,+}H^{b}_{\Delta_2,+}\sim &\frac{ig^2f^{abc}}{z_{12}}\sum_{m=0}^{\infty} \frac{\zb_{12}^m}{m!}
 \left[ G_1 +G_2\, \partial_{\D}+G_3\partial_{\D}^2 \right]  \bar{\partial}^m H^c_{\D, +}\Big|_{\D = \D_1+\D_2-1}  ,
\end{align}
where
\begin{align}
    &G_1 = \left(1
      -\frac{a \,\pi^2}{24}-\frac{a}{4}\log^2\left|z_{12}\right|^2 -a\left[\alpha_{0,2}+\alpha_{2,0}+\left(\alpha_{1,0}+\alpha_{0,1}\right)\log\left|z_{12}\right|^2 + \alpha_{1,1} \right]\right) C_+^{(0)}\nonumber\\
    &G_2 = -2a\,  C_+^{(0)} \left( \alpha_{1,0} + \alpha_{0,1} + \frac{1}{2} \log \left|z_{12}\right|^2 \right), \qquad G_3 = -a \, C_+^{(0)}\label{eq:hardgdef}
\end{align}
and the $\alpha_{i,j}$ are the integrals
\begin{align}
    \alpha_{i,j} = \frac{1}{C_+^{(0)}} \int_0^1 dt\,  t^{\D_1-2+m} (1-t)^{\D_2-2} \, \log^i t  \log^j (1-t) .
\end{align}
This integral can be evaluated explicitly expressions for all relevant cases.
\begin{align}
\label{eq:alphaints}
    \alpha_{2,0} &= \alpha_{1,0}^2 + \psi^{(1)} \left(\D_1+m-1\right) - \psi^{(1)} \left(\D_1+\D_2+m-2\right)\\
    \alpha_{0,2} &= \alpha_{0,1}^2 + \psi^{(1)} \left(\D_2-1\right) - \psi^{(1)} \left(\D_1+\D_2+m-2\right) \nonumber\\
    \alpha_{1,0} &= \psi^{(0)}\left(\D_1+m-1\right) - \psi^{(0)}\left(\D_1+\D_2+m-2\right)\nonumber\\
    \alpha_{0,1} &= \psi^{(0)}\left(\D_2-1\right) - \psi^{(0)}\left(\D_1+\D_2+m-2\right) \nonumber
\end{align}
Here,
\begin{equation}
    \psi^{(k)}(x) = \frac{d^k}{dx^k} \text{log} \left(\Gamma (x)\right),
\end{equation}
are the polygamma functions. We reemphasize that the finiteness we are referring to is only the behaviour as $\epsilon \to 0$. Indeed, (\ref{eq:hardOPEfinal}) is manifestly singular as the external particles become collinear $z_{12} \to 0$ and as we will see in the next section, it also exhibits singularities corresponding to external particles becoming soft.


\section{Conformally soft operators at one-loop}
\label{sec:confsoftdef}
Conformally soft operators capture the soft behaviour of amplitudes. The OPEs of such operators with hard operators capture key features of soft theorems and the OPEs with other conformally soft operators reveal symmetry algebras of the underlying theory. This is well studied at tree-level and before attempting to generalize these operators to loop level, it is instructive to first review various aspects of such operators at tree-level. Let us start with motivating their definition from the OPE (\ref{eq:hardOPEfinal}), which when restricted to tree-level reads  
\begin{equation}
\label{eq:treeOPEl}
    H^{a}_{\Delta_1,+}(z_1,\zb_1)H^{b}_{\Delta_2,+}(z_2,\zb_2)\sim\frac{ig^2 f^{abc}}{z_{12}} \sum_{m=0}^{\infty} \frac{\zb_{12}^m}{m!} \text{B}\left(\D_1-1+m, \D_2-1 \right) \, \bar{\partial}^m H_{\Delta_1+\Delta_2-1,+}^{c}.
\end{equation} 
Focusing on the singularity structure of OPE in $\D_1$, the Beta function on the RHS of the OPE has poles at all integer values of $\D_1$ less than $1$. Consistency of the OPE now implies that all tree-level correlators in Yang-Mills theory must have these poles. This is guaranteed by the soft theorems and this motivates the following definition of conformally soft operators \cite{Guevara:2021abz,Himwich:2021dau}(valid only at tree-level)
\begin{equation}
    \label{eq:treesoftops}
    R^{(k,1), a}  \equiv \lim_{\e \to 0} \varepsilon\,  H^a_{k+\varepsilon,+}.
\end{equation}
This $\varepsilon$ is distinct from $\epsilon$ used in the previous section as a dimensional regulator. The extra superscript ``1'' is meant to indicate that this operator corresponds to a simple pole. We follow this logic and define conformally soft operators at one-loop by looking at the singularity structure of the (\ref{eq:hardOPEfinal}). First, we note that the RHS has double and triple poles in $\D_1$ due to the terms involving single and double derivatives of the Euler Beta function. Thus consistency once again requires that all 1-loop correlators in Yang-Mills theory have double and triple poles and we can define a family of \emph{three} conformally soft operators
\begin{align}
    R^{(k,1)a}(z,\bar{z}) = \frac{1}{2}\lim_{\varepsilon\to 0}\partial^{2}_{\varepsilon}(\varepsilon^{3}H^{a}_{k+\varepsilon,+}(z,\bar{z})) &\qquad  R^{(k,2)a}(z,\bar{z}) = \lim_{\varepsilon\to 0}\partial_{\varepsilon}(\varepsilon^{3}H^{a}_{k+\varepsilon,+}(z,\bar{z})) \nonumber\\
    & \hspace{-2cm} R^{(k,3)a}(z,\bar{z}) = \lim_{\varepsilon\to 0}(\varepsilon^{3}H^{a}_{k+\varepsilon,+}(z,\bar{z})) \label{eq:1loopconfsoftdef}
\end{align}
These definitions capture the expectation that the operator can be schematically written as 
\begin{equation}
\label{eq:epsilonpoles}
    H_{k+\varepsilon, +}^a \sim \frac{R^{(k,3),a}}{\varepsilon^3} +\frac{R^{(k,2),a}}{\varepsilon^2} +\frac{R^{(k,1),a}}{\varepsilon}.
\end{equation}
Furthermore, $R^{(k,1)a}(z,\bar{z})$ reduces to the tree-level soft current in the absence of loop-corrections while $R^{(k,2)a}(z,\bar{z}), R^{(k,3)a}(z,\bar{z})$ vanish at tree-level since correlators only have simple poles in $\D_1$. Thus
\begin{align}
\label{eq:softopsadep}
      & R^{(k,1)a}(z,\bar{z}) \sim R^{(k,1)a}(z,\bar{z}) \vert_{\text{tree}} + a R^{(k,1)a}(z,\bar{z}) \vert_{\text{1-loop}},\nonumber \\
      & R^{(k,2)a}(z,\bar{z}) \sim \mO(a), \qquad  R^{(k,3)a}(z,\bar{z}) \sim \mO(a).
\end{align}
As always, the above equations are meant to be understood as statements about all correlators involving these operators. This schematic will prove to be useful in \ref{sec:softsoftOPE}.\\

We conclude this section with a discussion of how IR divergences can be included in conformally soft operators. The definition of conformally soft operators (\ref{eq:1loopconfsoftdef}) is based on the IR finite operators $H_{\D,+}^a$. Naturally, any OPEs involving them will be free of IR divergences - a fact that will be shown explicitly in the next section. An ostensibly inequivalent definition of these operators based on the bare operators $\mO_{\D,+}^a$ is
\begin{equation}
    \label{eq:baresoftops}
    \tilde{R}^{(k',3)a} \left(z,\zb\right) \equiv \lim_{\varepsilon \to 0} \varepsilon^3 \mO^a_{k'+\epsilon, +}\left(z,\zb\right),
\end{equation}
 with $\tilde{R}^{(k',1)a},\tilde{R}^{(k',2)a}$ defined analogously. However, such operators are simply related to the ones in (\ref{eq:softopsadep}) by 
 \begin{equation}
    \label{eq:barefiniterel}
    \tilde{R}^{(k',3)a} \left(z,\zb\right) = {\bf V}_{\kappa}\left(z,\zb\right) R^{(k,3)a} \left(z,\zb\right), \qquad k' = k - \frac{a}{2}C_{1,+}^{(0)},
\end{equation}
as expected from the decomposition (\ref{eq:decompoperator}). The relative shift in the location of the singularities can be explained as by recalling that the decomposition in (\ref{eq:decompoperator}) is based on the factorization of the amplitude into soft and hard factors which takes the schematic form 
\begin{equation}
    \label{eq:hardsoftfactamp}
    \mA_n = {\bf Z}_n \,\prod_{k=1}^n \om_k^{\frac{a}{2}C_{1,+}^{(0)}} \mathcal{H}_n.
\end{equation}
We refer the reader to \cite{Bhardwaj:2022anh} for more details. The singularities of the corresponding Celestial amplitude are thus located at $\Delta + \frac{a}{2}C_{1,+}^{(0)} = k$ where $k=1,0, -1, \dots$ justifying  (\ref{eq:barefiniterel}).
\section{OPEs of conformally soft operators at one-loop}
\label{sec:confsoftOPE}
Having defined conformally soft operators at one-loop (\ref{eq:1loopconfsoftdef}), we can now compute their OPEs with hard gluon operators and with themselves. At tree-level, the soft operators satisfy current algebras as shown in \cite{Pate:2019lpp, Guevara:2021abz}. They can thus be thought of as currents and knowing the OPE of these currents with the hard operators is equivalent to knowing how they transform under these symmetries. The Ward identities corresponding to these symmetries are equivalent to the soft theorems. All of these statements are expected to receive loop corrections. In particular, it is interesting to understand how the symmetry algebra observed at tree-level is modified by loop corrections. The first step in answering this question is to compute the one-loop corrections to the OPEs of soft operators. We will first compute the OPE of one hard and one soft operator before moving on to the OPE of two soft operators. 
\subsection{Hard-soft OPEs}\label{sec:hardsoftOPE}
It is helpful to start by outlining how the OPE of a conformally soft gluon with a hard one can be obtained at tree-level. This is done by taking the limit in (\ref{eq:treesoftops}) on (\ref{eq:treeOPEl}). Equivalently, we simply expand both sides in $\varepsilon$ and equate the coefficients of $\frac{1}{\varepsilon}$, resulting in
\begin{align}
\label{eq:softhardtree}
    R^{k,a}(z_1,\bar{z}_1)H_{\Delta_2,\pm}^{b}(z_2,\bar{z}_2) &= \lim_{\varepsilon\to 0}\varepsilon H_{k+\varepsilon,+}^{a}(z_1,\bar{z}_1)H^{b}_{\D_2,\pm}(z_2,\bar{z}_2)\\
    &\sim -g^2\frac{if^{abc}}{z_{12}}\sum_{m=0}^{1-k}\frac{\zb_{12}^m}{m!} \, \mathcal{G}_0 \left(m\right) \, \partial_{\zb_2}^mH_{\Delta_2+k-1,\pm}^{c}(z_2,\bar{z}_2)\nonumber,
\end{align}
where  
\begin{equation}
    \label{eq:treelevelsofthardcoeff}
\mathcal{G}_0 \equiv \lim_{\varepsilon \to 0} \varepsilon \, C_+^{(0)} = \frac{\Gamma(3-k-m-\Delta_2)}{\Gamma(2-k-m)\Gamma(2-\Delta_2)}.
\end{equation}
This procedure is now easily generalized to all the terms in the one-loop OPE (\ref{eq:IRfiniteOPE}) which has, in addition to the simple pole in $\varepsilon$, double and triple poles. The OPE of $R^{(k,i)a}$ with a hard gluon is extracted by equating the coefficient of $\frac{1}{\varepsilon^i}$ on both sides. This gives
\begin{align}
    &R^{(k,1)a}H_{\Delta_2,+}^{b}~\sim  g^2\frac{if^{abc}}{z_{12}}\sum_{m=0}^{1-k}\frac{\bar{z}^m_{12}}{m!}\, \bigg[\mathcal{G}^{(1)}_1 +\mathcal{G}_2^{(1)}\partial_{\Delta_2}+\mathcal{G}_3^{(1)}\partial^2_{\Delta_2}\bigg]\partial_{\bar{z_2}}^m H_{\Delta_2+k-1,+}^{c} \nonumber\\
     &R^{(k,2)a}H_{\Delta_2,+}^{b} \sim  g^2\frac{if^{abc}}{z_{12}}\sum_{m=0}^{1-k}\frac{\zb_{12}^m}{m!}\, \bigg[\mathcal{G}_1^{(2)} + \mathcal{G}_2^{(2)}\partial_{\Delta_2}\bigg]\partial_{\zb_2}^m H^{c}_{\Delta_2+k-1,+}\nonumber\\
     &R^{(k,3)a}H_{\Delta_2,+}^{b} \sim g^2\frac{if^{abc}}{z_{12}}\sum_{m=0}^{1-k}\frac{\zb_{12}^m}{m!} \mathcal{G}_1^{(3)} \,\partial_{\zb_2}^m H^{c}_{\Delta_2+k-1,+}. \label{eq:softhard1loop}
\end{align}
The $\mathcal{G}_i^{(j)}$ are the coefficients of $\frac{1}{\varepsilon^j}$ in the Laurent series expansion of the $G_i$ defined in (\ref{eq:hardgdef}). A direct computation yields
\begin{align}
   & \mathcal{G}^{(1)}_1 = \mathcal{G}_0 \,\left[1-a\left(\frac{1}{4}\log^2\left(|z_{12}|^2 \right)+\frac{\pi^2}{24}+\log \left(|z_{12}|^2 \right)\Psi^{(0)} +(\Psi^{(0)})^2 -\Psi^{(1)} \right)\right], \nonumber\\
   &\mathcal{G}^{(1)}_2 = -2a \, \mathcal{G}_0 \,\left[\Psi^{(0)}+\frac{1}{2}\log\left(\left|z_{12}\right|^2\right) \right], \qquad \mathcal{G}^{(2)}_1 =  a\, \mathcal{G}_0 \,\left( \log \left|z_{12}\right|^2 - \Psi^{(0)} \right)\nonumber \\
   & \hspace{4cm} \mathcal{G}^{(2)}_2 = -2\mathcal{G}_3^{(1)} =-\mathcal{G}_1^{(3)} = 2a \, \mathcal{G}_0. \label{eq:softgdef}
\end{align}
the following combination of Polygamma functions
\begin{equation}
   \Psi^{(n)} = \psi^{(n)}(2-\Delta_2)-\psi^{(n)}(3-k-m-\Delta_2), \label{eq:psidef}
\end{equation}
appears ubiquitously in the above equations. An explicit formula can be given for the OPE coefficients since $k,m$ are integers: 
\begin{align}
    &\mathcal{G}_0 \, \Psi^{(0)} = -\frac{1}{\Gamma\left(2-k-m\right)} \sum_{r=0}^{-(k+m)}\prod_{s=0,s\neq r}^{s=-(k+m)}\left(2-\D_2+s\right)
\end{align} 
Finally, we can include the effects of IR divergences by considering the OPEs of the operators (\ref{eq:barefiniterel}) which are closely related to the OPEs (\ref{eq:softhard1loop}). The IR divergent terms (terms which diverge as the dimensional regularization parameter $\epsilon \to 0$) are displayed in the equation below. The finite terms in this OPE are identical to that in (\ref{eq:softhard1loop}) and we have suppressed them 
\begin{align}
    &\tilde{R}^{(k,1)a}\mathcal{O}_{\Delta_2,+}^{b}\sim \frac{iag^2f^{abc}}{2z_{12}}\sum_{m=0}^{1-k}\frac{\bar{z}_{12}^m}{m!}\mathcal{G}_0 \bigg[-\frac{1}{\epsilon^2}+\frac{1}{\epsilon}\bigg(c_{E}+2\Psi^{(0)}+\log|z_{12}|^2+\partial_{\Delta_2}\bigg)\bigg]\partial^m_{\bar{z}_2}\mathcal{O}^{c}_{\Delta_2,+} + \mathcal{O}(\epsilon^0)\nonumber\\
    &\tilde{R}^{(k,2)a}\mathcal{O}_{\Delta_2,+}^{b}\sim -\frac{ag^2}{\epsilon}\frac{if^{abc}}{z_{12}}\sum_{m=0}^{1-k}\frac{\bar{z}_{12}^m}{m!}\mathcal{G}_0 \,\partial^m_{\bar{z}_2}\mathcal{O}^{c}_{\Delta_2,+}
    +\mathcal{O}(\epsilon^0)\nonumber\\
    &\tilde{R}^{(k,3)a}\mathcal{O}_{\Delta_2,+}^{b}\sim \mathcal{O}(\epsilon^0) \label{eq:divOPEterms}.
\end{align}

\subsection{Soft-soft OPEs}\label{sec:softsoftOPE}
 The second set of OPEs that we can extract from (\ref{eq:hardOPEfinal}) are the ones between two conformally soft operators. Multiple soft insertions are generally ill defined since soft limits do not commute \footnote{There are some exceptions to this statement for currents with $k,l = 1, 0$}. It is helpful to first briefly review how this can be a potential obstruction to defining the OPE of two soft currents even at tree-level. This argument was presented in \cite{Ball:2022bgg} where the author considered various ways of approaching the soft limit of (\ref{eq:treeOPEl}) and defined the operators 
 \begin{equation}
     \label{eq:doublesoftdef}
     R^{(l,1), a} \equiv \, \lim_{\varepsilon \to 0} \varepsilon \,\eta_1 H_{l + \eta_1 \, \varepsilon,+}^a, \qquad R^{(k,1), b} \equiv \lim_{\varepsilon \to 0} \varepsilon \,\eta_2 H_{k + \eta_2  \, \varepsilon}
 \end{equation}
 where $k,l = 1, 0, -1, \dots$. Here, $\eta_1, \eta_2$ are the variables which parametrize the rates of approaching the conformally soft limit. The OPE (at tree-level) of these two operators is
\begin{align}
    \label{eq:treelevelsoftsoft}
    &R^{(l,1), a}R^{(k,1), b} \equiv \lim_{\varepsilon\to 0} \varepsilon^2 \eta_1 \eta_2 H^{a}_{l+\eta_1 \varepsilon,+}(z_1,\zb_1)H^{b}_{k+\eta_2 \varepsilon,+}(z_2,\zb_2)\\
    &\,\,\sim\frac{ig^2 f^{abc}}{z_{12}} \sum_{m=0}^{1-l} \left[\binom{2-k-l-m}{1-k}\, + \frac{(-1)^{1-k}}{1+\frac{\eta_1}{\eta_2}}\sum_{m=3-k-l}^{\infty} \binom{l-2+m}{1-k}\right]\frac{\zb_{12}^m}{m!} \bar{\partial}^m R^{(k+l-1,1) c}\nonumber.
\end{align}
A striking feature of this formula is the dependence of the OPE coefficients on $\eta_1, \eta_2$ for $m\geq 3-k-l$. This indicates that the OPE coefficients are not independent of the way in which the conformally soft limit is approached. Consequently, the OPE coefficients would not be well defined. However, the soft current $R^{(k+l-1,1)c}\left(z_2, \zb_2 \right)$ is a polynomial of degree $1-k-l$ in the collinear limit considered here. All terms with $m>3-k-l$ thus vanish since the derivative annihilates the soft operator. The OPEs of two soft operators are thus well defined. \\
We can now include loop corrections and repeat the above analysis. At first glance, there seem to be $9$ different OPEs between the three conformally soft operators $R^{(k,1),a}, R^{(k,2),a}, R^{(k,3),a}$ corresponding to the 9 possible pairs $\left\lbrace R^{(k,m),a},R^{(l,n),a} \right\rbrace$. However, double and triple poles are loop effects and any correlator with higher order poles in two conformal dimensions must necessarily be $\mathcal{O}(a^2)$, where $a$ is the `t Hooft coupling. These cannot be reliably computed solely from one-loop OPEs which only involve $\mathcal{O}(a)$ corrections. This is also made clear from the schematic (\ref{eq:softopsadep}). Thus, the only relevant OPEs are 
\begin{align}
    R^{(l,m)a}R^{(k,1)b}&\equiv \frac{\eta_1^{m} \eta_2}{2(3-m)!}\lim_{\varepsilon\to 0}\partial^{3-m}_{\varepsilon}\left(\varepsilon^{3}H^{a}_{l+\eta_1 \varepsilon,+}\right) \, \partial^{2}_{\varepsilon}\left(\varepsilon^{3}H^{b}_{k+\eta_2 \varepsilon,+}\right)\label{eq:soft_current_OPE_loop}, \qquad m=1,2,3. 
\end{align} 
\paragraph{${\bf R^{(l,1)}R^{(k,1)}}$ OPE:} We can obtain this OPE simply by equating the coefficients of $\frac{1}{\varepsilon^2}$ in the Laurent expansion of both sides of (\ref{eq:hardOPEfinal}). We will present results only for $(k,l) = (1,1), (1,0), (0,0)$ which serves to demonstrate the dependence of these OPEs on the order of limits. The relevant OPEs are as follows
\begin{align}
    R^{(1,1),a}R^{(1,1),b} &\sim\nn\frac{ig^2f^{abc}}{z_{12}}\bigg[R^{(1,1),c}+a\bigg\{\bigg(\frac{\pi^2}{12}+\frac{10\pi^2\eta_1^3\eta_2^3}{(\eta_1+\eta_2)^6}+\frac{1}{4}\log^2(z_{12}\zb_{12})\bigg)R^{(1,1),c}\\
    &\qquad-\frac{5\pi^2\eta_1^3\eta_2^3}{(\eta_1+\eta_2)^3}R^{(1,3),c}\bigg\}\bigg]~, \\
    R^{(0,1),a}R^{(1,1),b}&\sim \frac{ig^2f^{abc}}{z_{12}}\bigg[R^{(0,1),c}+a\bigg\{\bigg(\frac{\pi^2}{12}+\frac{10\pi^2\eta_1^3\eta_2^3}{(\eta_1+\eta_2)^6}-\frac{\eta_1^3(\eta_1^2+5\eta_1\eta_2+10\eta_2^2)}{2(\eta_1+\eta_2)^5}\log{(z_{12}\zb_{12})}\nn\\
    &\qquad+\frac{1}{4}\log^2{(z_{12}\zb_{12})}\bigg)R^{(0,1),c}
    -\frac{\eta_1^3(\eta_1^2+5\eta_1\eta_2+10\eta_2^2)}{2(\eta_1+\eta_2)^5}R^{(0,2),c}\nn\\
    &\qquad-\frac{5(6+\pi^2)\eta_1^3\eta_2^3}{(\eta_1+\eta_2)^6}R^{(0,3),c}\bigg\}\bigg]~, \\
    R^{(0,1),a}R^{(0,1),b} &\sim \frac{ig^2f^{abc}}{z_{12}} (2+\zb_{12}\bar{\partial})\bigg[R^{(-1,1),c}+a\bigg\{\bigg(\frac{\pi^2}{12}+\frac{10\pi^2\eta_1^3\eta_2^3}{(\eta_1+\eta_2)^6}-\frac{1}{4}\log(z_{12}\zb_{12})\nn\\
    &\qquad+\frac{1}{4}\log^2(z_{12}\zb_{12})\bigg)R^{(-1,1),c}-R^{(-1,2),c}-\frac{5(6+\pi^2)\eta_1^3\eta_2^3}{(\eta_1+\eta_2)^6}R^{(-1,3),c}\bigg\}\bigg]~.
\end{align}

\section{Soft theorems and OPEs}
\label{sec:OPEsofttheorem}
\subsection{Tree-level}
The tree-level OPE (\ref{eq:treeOPEl}) suggests that all correlators have simple poles at $\D=1, 0, -1, \dots$. To see that this is indeed the case, we consider the quantity $\lim_{\D_1 \to k} \left(\D_1 - k \right) \mAt_n$. A non zero result would confirm the presence of a pole. Using the definition of the celestial amplitude (\ref{eq:celampdef}) and the identity
\begin{equation}
\label{eq:deltaidentity}
    \lim_{\D_1 \to k} \left(\D_1 - k\right) \om_1^{\D_1 -1} = \delta \left(\om_1 \right) \om_1^{k-1},
\end{equation}
we arrive at
\begin{equation}
\label{eq:softint1}
    \lim_{\D_1 \to k} \left(\D_1 - k \right) \mAt_n = \int d\om_1 \, \delta \left(\om_1 \right) \om_1^{k-1}\, \int \prod_{i=2}^n d\om_i\,  \om_i^{\D_i-1} \mA_n . 
\end{equation}
The $\om_1$ integral is now completely determined by the soft behaviour (in $\om_1$) of the scattering amplitudes which is given by 
\begin{equation}
    \mathcal{A}_{n+1} \xrightarrow[]{\om_1 \to 0} \left( \frac{1}{\omega_1} S^{(0)} + \omega_1^0 S^{(1)}  \right) \mathcal{A}_n + \sum_{i=1}^{\infty} S^{(i)} \om_1^{i} B^{(i)}_n.
    \label{soft theorem}
\end{equation} 
For Yang-Mills, the leading and subleading soft terms in this expansion are universal \cite{Weinberg:1965nx, Casali:2014xpa}. Rewritten in the notation of this paper, they are
\begin{align}
\label{eq:universalsfactorsmom}
    S_{+}^{(0)} &= \frac{1}{2}\frac{z_{n2}}{z_{12}z_{n1}}, \\
    \quad S_{+}^{(1)} &= \frac{1}{2}\left[\frac{1}{ \om_2 z_{12}}\left(\sigma_2+\om_2 \frac{\partial}{\partial \om_2}+\zb_{12}\frac{\partial}{\partial \zb_2}\right) +\frac{1}{ \om_n z_{1n}}\left(\sigma_n+\om_n \frac{\partial}{\partial \om_n}+\zb_{1n}\frac{\partial}{\partial \zb_n}\right)\right].
\end{align}
These factors are to be understood as acting on the colour ordered amplitudes. The trace decomposition of the amplitude (\ref{eq:tracedecomptree}) implies that the soft factor acting on the full amplitude takes the form
\begin{align}
    \label{eq:coloureduniversalsfactorsmom}
    S_+^{(0)a_1a_ie} &= \frac{if^{a_1a_ie}}{2z_{1i}}~,\quad
    S_+^{(1)a_1a_ie} = \frac{if^{a_1a_ie}}{2z_{1i}\,\om_i}\left(\sigma_i+\om_i\frac{\partial}{\partial \om_i}+\zb_{1i}\frac{\partial}{\partial \zb_{2}}\right) ~.
\end{align}
the above coloured soft-factors can be worked out in the DDM basis \cite{DelDuca:1999rs} in which any tree-level amplitude can be written as 
\begin{equation}
    A_n^{a_1\cdots a_n,(0)} = ag^{n-2} \sum_{s \in S_{n-2}}f^{a_1a_{s(2)}b_1}f^{b_1a_{s(3)}b_2}\cdots f^{b_{n-3}a_{s(n-1)}a_n}A_n^{(0)}[1 s(2) s(3)\cdots s(n-1)n]~.
\end{equation}
Note that while $S^{(0)}$ is just a multiplicative factor, $S^{(1)}$ is a differential operator acting on the $n$ point amplitude. The rest of the terms in this expansion are not universal, i.e. the $B^{(i)}_n$ are theory specific functions that cannot be expressed in terms of the lower point amplitudes. However, for MHV amplitudes (and for the collinear part of any amplitude), all terms are universal and we can write  
\begin{equation}
    \mathcal{A}^{\text{MHV}}_{n+1} = \left( \sum_{k=-\infty}^{1} \omega_1^{-k} S^{(k)}\right) \mathcal{A}^{\text{MHV}}_n,
     \label{eq:softexp}
\end{equation}
where the $S^{(k)}$ do not depend on $\om_1$. We refer to \cite{Mago:2021wje} for an explicit expression of these factors, to \cite{Cachazo:2008hp} for a slightly more elaborate discussion on the meaning of ``collinear part''. For the rest of this paper, we will focus on MHV amplitudes and drop the superscript with the understanding that all the statements made here can be extended to the collinear parts of any amplitude beyond MHV. We are now in a position to perform the $\om_1$ integral in (\ref{eq:softint1}), which yields
\begin{align}
     \lim_{\D_1 \to k} \left(\D_1 - k \right) \mAt_n = \tilde{S}^{(k)} \mAt_{n-1}  \neq 0,
\end{align}
thereby demonstrating the presence of simple poles in correlators at $\D_1 = k$. The $\tilde{S}^{(k)}$ are now operators acting on celestial amplitudes. The operators acting on colour ordered celestial amplitudes can be inferred from (\ref{eq:universalsfactorsmom}) to be
\begin{align}
\label{eq:universalsfactorcel}
    &\tilde{S}^{(0)}_+ = \frac{1}{2}\frac{z_{n2}}{z_{12}z_{n1}}, \\
    &\nonumber \tilde{S}^{(1)}_+ = \frac{1}{2}\left[\frac{1}{z_{12}}\left(\sigma_2+1-\D_2+\zb_{12}\frac{\partial}{\partial \zb_2}\right)e^{-\partial_{\D_2}} + \frac{1}{ z_{1n}}\left(\sigma_n+1-\D_n+\zb_{1n}\frac{\partial}{\partial \zb_n}\right)e^{-\partial_{\D_n}}\right].
\end{align}
and from (\ref{eq:universalsfactorcel}) the coloured soft factors are given as
\begin{align}
\label{eq:coloureduniversalsfactorcel}
    \tilde{S}^{(0)a_1a_{i}e}_+ = \frac{if^{a_1a_{i}e}}{2z_{1i}}, \quad \tilde{S}^{(1)a_1a_{i}e}_+ = \frac{if^{a_1a_{i}e}}{2z_{1i}}\left(\sigma_i+1-\D_i+\zb_{1i}\frac{\partial}{\partial \zb_i}\right) e^{-\partial_{\D_i}}.
\end{align}
$\forall~i\in \{2,\dots,n\}$. It can be explicitly seen here that these satisfy 
\begin{align}
    \label{eq:nullstate}
    \bar{\partial}_1S_+^{(0)} = \bar{\partial}_1^2 S_+^{(1)} = 0. 
\end{align}
Finally, note that the leading (and subleading) soft limit commutes with the collinear limit. This is easy to see since we recover the soft theorems by applying the soft-hard OPE (\ref{eq:softhardtree}) within a correlator, i.e
\begin{align}
    &\Big< R^{1,a_1}(z_1,\bar{z}_1)H_{\Delta_2,\pm}^{a_2}(z_2,\bar{z}_2) \dots \Big> \sim -g^2\frac{if^{a_1a_2c}}{z_{12}}\Big< H_{\Delta_2,\pm}^{c}(z_2,\bar{z}_2) \dots \Big>\\
     &\Big< R^{0,a_1}(z_1,\bar{z}_1)H_{\Delta_2,\pm}^{a_2}(z_2,\bar{z}_2) \dots \Big> \sim -g^2\frac{if^{a_1a_2c}}{z_{12}}\left(1\pm 1-\D_2+\zb_{12}\partial_{\zb_2}\right) \Big< H_{\Delta_2-1,\pm}^{c}(z_2,\bar{z}_2) \dots \Big>,\nonumber
\end{align}
which agrees with (\ref{eq:coloureduniversalsfactorcel}).
\subsection{One-loop}
At one-loop level in Yang-Mills theory, even the leading soft theorem is corrected. This was first derived in \cite{Bern:2014oka}. Adapted to the notations of this paper, it reads
\begin{equation}
\label{eq:1loopsoftthm}
    \mA_{n+1}^{(1)} \xrightarrow[]{\om_1 \to 0} S^{(0)}_+ \mA_n^{(1)} + S^{(0), 1-\text{loop}}_+ \mA_n^{(0)}
\end{equation}
where $S_+^{(0)}$ is the tree-level soft factor defined in (\ref{eq:universalsfactorcel}) and $S_+^{(0),1-\text{loop}}$ is the one-loop correction given by 
\begin{align}
    S^{(0), 1-\text{loop}}_+ &= -S^{(0)}_+\frac{c_{\Gamma}}{\epsilon^2}\left(\frac{-\mu^2s_{n2}}{s_{12}s_{n1}}\right)^{\epsilon}\frac{\pi \epsilon}{\sin{\pi \epsilon}} \label{eq:1loopsoftfactor}\\
    &= -S^{(0)}_{+}c_{\Gamma}\left(\frac{1}{\epsilon^2}+\frac{1}{\epsilon}\log{\left(\frac{-\mu^2s_{n2}}{s_{12}s_{n1}}\right)}+\frac{1}{2}\log^2{\left(\frac{-\mu^2s_{n2}}{s_{(12}s_{n1}}\right)}+\frac{\pi^2}{6}\right)+\mathcal{O}(\epsilon).\nonumber
\end{align}
Here $s_{ij} = \an{ij}\sq{ji}$. Rewriting the soft factor in the parametrization (\ref{eq:momentaparameterization}), it takes the form
\begin{align}
    \label{eq:1loopsoftcel}
     S^{(0), 1-\text{loop}}_+ = -\frac{a\, c_{\Gamma}}{2\om_1}\frac{z_{2n}}{z_{12}z_{1n}}&\Bigg[\frac{1}{\epsilon^2}+\frac{\pi^2}{6}+\frac{1}{\epsilon}\log \left(\frac{-\left|z_{n2}\right|^2}{\left|z_{12}\right|^2\left|z_{n1}\right|^2}\right)-\left(\frac{2}{\epsilon}+\log\left(\frac{-\left|z_{n2}\right|^2}{\left|z_{12}\right|^2\left|z_{n1}\right|^2}\right)\right)\log{\om_1}\nonumber \\
     &\quad +\frac{1}{2}\log^2\left(\frac{-\left|z_{n2}\right|^2}{\left|z_{12}\right|^2\left|z_{n1}\right|^2}\right)+2 \log^2 \om_1 \Bigg].
\end{align}
The conformally soft operators $R^{(k,1), a}, R^{(k, 2), a}$ and $R^{(k,3), a}$ in (\ref{eq:1loopconfsoftdef}) were defined such that they corresponded to extracting the coefficients of $\frac{1}{\om_1}, \frac{\log \om_1}{\om_1}$ and $\frac{\log^2 \om_1}{\om_1}$ respectively. We can thus extract the following three conformally soft theorems for colour ordered amplitudes from (\ref{eq:1loopsoftthm}, \ref{eq:1loopsoftfactor})
\begin{align}
    \label{eq:1loopcsoftthm}
   & \an{R^{(1,1)} \mO_{\D_2, +} \dots \mO_{\D_n, +}} = \an{\mO_{\D_2, +} \dots \mO_{\D_n, +}}\\
   \nonumber &\qquad   \times\frac{1}{2}\frac{z_{2n}}{z_{12}z_{1n}} \left[1-a\, c_{\Gamma}\left(\frac{\pi^2}{6} + \frac{1}{\epsilon^2}+\frac{1}{\epsilon}\log \left(\frac{-\left|z_{n2}\right|^2}{\left|z_{12}\right|^2\left|z_{n1}\right|^2}\right) + \frac{1}{2} \log^2 \left(\frac{-\left|z_{n2}\right|^2}{\left|z_{12}\right|^2\left|z_{n1}\right|^2}\right)\right)\right]~,\\
   &\nonumber \an{R^{(1,2)} \mO_{\D_2, +} \dots \mO_{\D_n, +}} = \frac{-a\, c_{\Gamma}}{2}\frac{z_{2n}}{z_{12}z_{1n}}\log \left(\frac{-\left|z_{n2}\right|^2}{\left|z_{12}\right|^2\left|z_{n1}\right|^2}\right)  \an{ \mO_{\D_2, +} \dots \mO_{\D_n, +}}~,\\
   &\nonumber \an{R^{(1,3)} \mO_{\D_2, +} \dots \mO_{\D_n, +}} = -a g^2\, c_{\Gamma}\frac{z_{2n}}{z_{12}z_{1n}} \an{ \mO_{\D_2, +} \dots \mO_{\D_n, +}}~.
\end{align}
These are the conformally soft theorems at one-loop for Yang-Mills theory. We conclude this section by noting some properties of the soft limits and currents. Firstly, the leading soft limit and the collinear limit fail to commute. This can be seen by computing the holomorphic collinear limit of (\ref{eq:1loopcsoftthm}) and comparing the result with the OPEs in  (\ref{eq:divOPEterms}, \ref{eq:softhard1loop}). Secondly, these currents are clearly not polynomials in $\zb_1$ and consequently do not satisfy equations similar to (\ref{eq:nullstate}).
\section{Loop level soft currents as logarithmic descendants}
\label{sec:logcft}
The OPE in (\ref{eq:hardOPEfinal}) involves terms which depend on $\log \left|z_{12}\right|^2$, with $|z_{12}|^2$ being the separation of the two operators on the celestial sphere. It was already pointed out in \cite{Bhardwaj:2022anh} that this raised the possibility of the CCFT being similar to a logarithmic CFT. In fact, it has also been noted that such structures arise even at tree-level in theories with dynamical gravity \cite{Fiorucci:2023lpb}. Such CFT, which was first discussed in \cite{Gurarie:1993xq} are characterized by the property that the Dilatation operator is non-diagonalizable. Consequently, states are organised in logarithmic multiplets. A multiplet of rank $r\geq 1$~\cite{Hogervorst:2016itc} is built on the top of $r$ primary operators with identical conformal dimension. These operators mix under the action of the dilatation operator rendering it non-diagonalizable. In this section, we will demonstrate that the soft currents $R^{(k,1)},R^{(k,2)},R^{(k,3)}$ form a rank-2 logarithmic multiplet. In order to do this, we first briefly flesh out the transformation properties of various operators in a logarithmic CFT. \\
We will denote a logarithmic multiplet of rank $r$ involving primaries with conformal dimension $\D$ and spin $J$ by the boldfaced symbol $\mathbf{O}^r_{\D,J}\left(z, \zb\right)$. Elements of this multiplet will be denoted by $O^{(a)}_{\D,J}\left(z, \zb\right)$ with $a=1, \dots r+1$. Under a conformal transformation, 
\begin{equation}
z \to z' = f(z), \qquad \zb \to \zb' = \bar{f}(\zb),    
\end{equation}
an element of this multiplet transforms as
\begin{equation}
    O^{(a)}_{\D,J}\left(z, \zb\right) \to O'^{(a)}_{\D,J}\left(z', \zb'\right) = \left(\partial z'\right)^{\Delta+J}\left(\bar{\partial} \zb'\right)^{\Delta-J}\sum_{b=a}^{r}\frac{\log^{b-a}\left|\partial z' \right|^2}{(b-a)!}O^{(b)}_{\D,J}\left(z, \zb\right)\,.\label{eq:logmultiplettransform}
\end{equation}
A natural way of constructing such a multiplet is by appending to a primary operator, derivatives w.r.t its conformal dimension~\cite{Hogervorst:2016itc, Flohr:2001zs}. Explicitly, $O^{(a)}_{\D,J} = \frac{1}{(r-a)!} \partial_{\D}^{r-a} O_{\D,J}^{(1)}$ and
\begin{equation}
    \mathbf{O}^{r}_{\D,J} \equiv \left\lbrace O^{(1)}_{\D,J}, \partial_{\D} O^{(1)}_{\D,J}, \dots ,\frac{1}{(r-a)!}\partial^{r-a}_{\Delta}O^{(1)}_{\Delta,J}, \dots ,\frac{1}{r!}\partial^{r}_{\Delta}O_{\Delta,J}  \right\rbrace.
\end{equation}
Returning now to CCFT, we recall that the soft currents are defined by 
\begin{equation}
    R^{(k,m)a}(z,\bar{z}) = \frac{1}{(3-m)!}\lim_{\varepsilon \to 0} \partial^{3-m}_{\varepsilon} \left(\varepsilon^{3}H_{k+\varepsilon,+}^{a}(z,\bar{z})\right), \quad m=1, 2, 3. 
\end{equation}
The transformation properties of these operators follow from those of $H_{\D,+}^a$ which transforms as a primary conformal dimension $\D$ and spin 1. This in turn follows from (\ref{eq:decompoperator}) after recalling that the operator ${\bf V}_{\kappa} \left(z,\zb\right)$ is a vertex operator which transforms as a primary with weight $-\frac{a}{2}C_{1,+}^{(1)}$ (c.f.(\ref{eq:++OPEfinalcoeffs}). Note that here, we have exchanged the labels $\D$ and $\D'$ in line with the previous sections of this paper. Thus
\begin{align*}
    R^{(k,m)a}(z,\bar{z}) &\to \frac{1}{(3-m)!}\lim_{\varepsilon\to 0}\partial^{3-m}_{\varepsilon}\left(\left(\partial z'\right)^{k+\varepsilon+1}\left(\partial\bar{z}'\right)^{k+\varepsilon-1}\varepsilon^{3}H^{a}_{k+\varepsilon}(z,\zb)\right)\\
    &= \frac{1}{(3-m)!}\lim_{\varepsilon\to 0}\sum_{r=0}^{3-m} {3-m\choose r} \partial^r_{\varepsilon}\left[\left(\partial z'\right)^{k+\varepsilon+1}\left(\partial\bar{z}'\right)^{k+\varepsilon-1}\right]\partial^{3-m-r}_{\varepsilon}\left(\varepsilon^{3}H_{k+\varepsilon,+}\left(z,\zb\right)\right)\\
    &= \left(\partial z'\right)^{k+1}\left(\partial\bar{z}'\right)^{k-1}\sum_{r=0}^{3-m}\frac{1}{r!}\log^r \left|\partial z'\right|^2\frac{1}{(3-m-r)!}\lim_{\varepsilon\to 0}\partial^{3-m-r}_{\varepsilon}\left(\varepsilon^{3}H_{k+\varepsilon}(z,zb)\right)\\
    &=\left(\partial z'\right)^{k+1}\left(\partial\bar{z}'\right)^{k-1}\sum_{r=0}^{3-m} \frac{1}{r!}\log^r\left|\partial z'\right|^2 \, R^{(k,m)a}(z,\zb)
\end{align*}
which is precisely the transformation law (\ref{eq:logmultiplettransform}), thus showing that the three soft currents form a logarithmic multiplet. 

\section{Discussion}
\label{sec:discussion}
In this paper, we have investigated a number of properties of soft currents. The one-loop soft currents seem to form logarithmic multiplets but fail to satisfy the naive null state equations of their tree-level counterparts. It would be interesting to see if the analysis of \cite{Pasterski:2021dqe, Pasterski:2021fjn, Pano:2023slc} can be extended to include null states logarithmic CFTs ~\cite{Flohr:2001zs, Nivesvivat:2020gdj,Hogervorst:2016itc,Rasmussen:2004jc} and if the soft theorems can be deduced from them. We have shown that simultaneous conformal soft limits are ill defined. However, it should be noted that this isn't necessarily an obstruction to understanding how the $\mathcal{S}-$algebra is modified at one-loop. The action of symmetries can be consistently defined by consecutive soft limits \cite{Ball:2022bgg}. The presence of logarithms complicates this analysis. For example, unlike the tree-level case it is not clear whether the conformally soft operators $R^{(k,m)}(z,\zb)$ can be mode expanded as a polynomial in $\zb$, the presence of logarithms in the OPE of primary operators perhaps could hinder the possibility of having a holomorphic mode expansion. Therefore, the usual CFT definition of a commutator of distinct local operators as a contour integral of an OPE does not apply. One would instead need to work with a more primitive definition of a commutator of two mode operators. This has been extensively discussed in \cite{Gurarie:2004ce} in the context of OPE of the stress tensor and its logarithmic partner. Also it is worth noting that in our work we speculate that at $\ell$-loop conformally soft currents form a rank $\ell+1$ logarithmic multiplet. At the same time, in section \ref{sec:softsoftOPE} we explicitly showed that these operators fail to define an OPE amongst themselves, thus tend to be non-local in nature. Such non-local logarithmic operators have previously been studied by Cardy in the context of certain critical limits of central charges in ordinary CFTs \cite{Cardy:2013rqg}. We hope to shed further light on both the extension to the $\mathcal{S}$-algebra beyond tree-level and the relationship to logarithmic CFTs sometime in the near future. 
\acknowledgments
We thank Hare Krishna and Adam Ball for useful discussion. We thank Marcus Spradlin and Anastasia Volovich for comments on the draft. AYS is supported by the STFC grant ST/X000761/1 and the Simons Collaboration on Celestial Holography, RB by the US Department of Energy under contract {DE}-{SC}0010010 (Task F)
\appendix

\section{Conformal invariance of (\ref{eq:loopOPEblock})}\label{app:Confinvarcheck}
We check the invariance of \eqref{eq:loopOPEblock} under infinitesimal $SL(2,\mathbb{R})_{L}\times SL(2,\mathbb{R})_{R}$ conformal transformations. As the left and the right sectors are independent of each other, we perform our analysis for the $SL(2,\mathbb{R})_{R}$ sector, the derivation for the  $SL(2,\mathbb{R})_{L}$ sector follows identically. 
\\

 An infinitesimal $SL(2,\mathbb{R})_{R}$ conformal transformation on an operator of conformal weight $(h,\bar{h})$ is defined as 
 \begin{equation}
     \delta_{\zb}\mathcal{O}_{h,\bar{h}}(z,\zb) = \left(\veb\hspace{0.5mm}\partial_{\zb}+\bar{h}\partial_{\zb}\veb\right)\mathcal{O}_{h,\bar{h}}(z,\zb)~.
 \end{equation}
The LHS of \eqref{eq:loopOPEblock} transforms as 
\begin{align}
\delta_{\zb}\left(\mathcal{O}_{\Delta_1,+}(z_1,\zb_1)\mathcal{O}_{\Delta_2,+}(z_2,\zb_2)\right) &= \delta_{\zb}\left(\mathcal{O}_{\Delta_1,+}(z_1,\zb_1)\right)\mathcal{O}_{\Delta_2,+}(z_2,\zb_2)+\mathcal{O}_{\Delta_1,+}(z_1,\zb_1)\delta_{\zb}\left(\mathcal{O}_{\Delta_2,+}(z_2,\zb_2)\right)\nonumber\\
&= \left(\veb_1\hspace{0.5mm}\partial_{\zb_1}+\frac{\Delta_1-1}{2}\partial_{\zb_1}\veb_1\right)\mathcal{O}_{\Delta_1,+}(z_1,\zb_1)\mathcal{O}_{\Delta_2,+}(z_2,\zb_2)+(1\leftrightarrow 2)
\end{align}
where $\veb_i \equiv \veb(\zb_i)$. The first term is given as 
\begin{align}
    &\veb_1\hspace{0.5mm}\partial_{\zb_1}\mathcal{O}_{\Delta_1,+}(z_1,\zb_1)\mathcal{O}_{\Delta_2,+}(z_2,\zb_2)+(1\leftrightarrow 2)
    \nn\\ 
    &\qquad\sim\frac{1}{z_{12}}\bigg[\left(1-\frac{\pi^2}{6}\right)\int_{0}^{1}dt~t^{\Delta_1-2}(1-t)^{\Delta_2-2}(t\veb_1+(1-t)\veb_2)\partial_{\zb}\mathcal{O}_{\Delta,+}(z_2,\zb)\nn\\
    &\qquad~~-\hat{c}_{\Gamma}\left(\frac{-\mu^2}{z_{12}\zb_{12}}\right)^{\e}\int_{0}^{1}dt~t^{\Delta_1-2-\e}(1-t)^{\Delta_2-2-\e}(t\veb_1+(1-t)\veb_2)\partial_{\zb}\mathcal{O}_{\Delta-2\e,+}(z_2,\zb)\nn\\
    &\qquad~~+2\hat{c}_{\Gamma}\partial_{\Delta_1}\partial_{\Delta_2}\int_{0}^{1}dt~t^{\Delta_1-2}(1-t)^{\Delta_2-2}(t\veb_1+(1-t)\veb_2)\partial_{\zb}\mathcal{O}_{\Delta,+}(z_2,\zb)\nn\\
    &\qquad~~-\hat{c}_{\Gamma}\frac{\e(\veb_1-\veb_2)}{\zb_{12}} \left(\frac{-\mu^2}{z_{12}\zb_{12}}\right)^{\e}\int_{0}^{1}dt~t^{\Delta_1-2-\e}(1-t)^{\Delta_2-2-\e}\mathcal{O}_{\Delta-2\e,+}(z_2,\zb)\bigg]\bigg|_{\Delta = \Delta_1+\Delta_2-1}
\end{align}
where $\zb=\zb_2+t \zb_{12}$, and the second term is 
\begin{align}
    &\frac{\Delta_1-1}{2}\partial_{\zb_1}\veb_1\mathcal{O}_{\Delta_1,+}(z_1,\zb_1)\mathcal{O}_{\Delta_2,+}(z_2,\zb_2)+(1\leftrightarrow 2) \nn\\
    &\qquad\sim \frac{1}{z_{12}}\bigg[\left(1-\frac{\pi^2}{6}\right)\int_{0}^{1}dt~t^{\Delta_1-2}(1-t)^{\Delta_2-2}(\hb_1\partial_{\zb_1}\veb_1+\hb_2\partial_{\zb_2}\veb_2)\mathcal{O}_{\Delta,+}(z_2,\zb)\nn\\
    &\qquad~~-\hat{c}_{\Gamma}\left(\frac{-\mu^2}{z_{12}\zb_{12}}\right)^{\e}\int_{0}^{1}dt~t^{\Delta_1-2-\e}(1-t)^{\Delta_2-2-\e}(\hb_1\partial_{\zb_1}\veb_1+\hb_2\partial_{\zb_2}\veb_2)\mathcal{O}_{\Delta-2\e,+}(z_2,\zb)\nn\\
      &\qquad~~+2\hat{c}_{\Gamma}\partial_{\Delta_1}\partial_{\Delta_2}\int_{0}^{1}dt~t^{\Delta_1-2}(1-t)^{\Delta_2-2}(\hb_1\partial_{\zb_1}\veb_1+\hb_2\partial_{\zb_2}\veb_2)\mathcal{O}_{\Delta,+}(z_2,\zb)\bigg]\bigg|_{\Delta = \Delta_1+\Delta_2-1}
\end{align}
where $\hb_i=\frac{\Delta_i-\sigma_i}{2}$. The RHS on the other hand transforms as  
\begin{align}
    &\frac{1}{z_{12}}\bigg[\left(1-\frac{\pi^2}{6}\right)\int_{0}^{1}dt~t^{\Delta_1-2}(1-t)^{\Delta_2-2}\left(\veb~\partial_{\zb}\mathcal{O}_{\Delta,+}(z_2,\zb)+(\hb_1+\hb_2)\partial_{\zb}\veb~\mathcal{O}_{\Delta,+}(z_2,\zb)\right)\nn\\
    &~-\hat{c}_{\Gamma}\left(\frac{-\mu^2}{z_{12}\zb_{12}}\right)^{\e}\int_{0}^{1}dt~t^{\Delta_1-2-\e}(1-t)^{\Delta_2-2-\e}\left(\veb~\partial_{\zb}\mathcal{O}_{\Delta-2\e,+}(z_2,\zb)+(\hb_1+\hb_2-\e)\partial_{\zb}\veb~\mathcal{O}_{\Delta-2\e,+}(z_2,\zb)\right)\nn\\
    &~+2\hat{c}_{\Gamma}\partial_{\Delta_1}\partial_{\Delta_2}\int_{0}^{1}dt~t^{\Delta_1-2}(1-t)^{\Delta_2-2}\left(\veb~\partial_{\zb}\mathcal{O}_{\Delta,+}(z_2,\zb)+(\hb_1+\hb_2)\partial_{\zb}\veb~\mathcal{O}_{\Delta,+}(z_2,\zb)\right)\bigg]\bigg|_{\Delta = \Delta_1+\Delta_2-1}
\end{align}
again note that $\veb \equiv \veb(\zb)$. Consider the difference between the RHS and the LHS of \eqref{eq:loopOPEblock} now, the result is 
\begin{equation}
    \delta_{\zb_2}(\text{RHS $-$ LHS of \eqref{eq:loopOPEblock}}) =   \circled{1} +\circled{2}+\circled{3}
\end{equation}
where we have 
\begin{align}
    \circled{1} &= \frac{1}{z_{12}}\left(1-\frac{\pi^2}{6}\right)\int_{0}^{1}dt~t^{\Delta_1-2}(1-t)^{\Delta_2-2}\bigg[\veb~\partial_{\zb}\mathcal{O}_{\Delta,+}(z_2,\zb)+(\hb_1+\hb_2)\partial_{\zb}\veb~\mathcal{O}_{\Delta,+}(z_2,\zb)\nn\\
    &~~-(\hb_1\partial_{\zb_1}\veb_1+\hb_2\partial_{\zb_2}\veb_2)\mathcal{O}_{\Delta,+}(z_2,\zb)-(t\veb_1+(1-t)\veb_2)\partial_{\zb}\mathcal{O}_{\Delta,+}(z_2,\zb)\bigg]\bigg|_{\Delta = \Delta_1+\Delta_2-1}\label{eq:term_1}\\
    \circled{2} &= \frac{\hat{c}_{\Gamma}}{z_{12}}\left(\frac{-\mu^2}{z_{12}\zb_{12}}\right)^{\e}\int_{0}^{1}dt~t^{\Delta_1-2-\e}(1-t)^{\Delta_2-2-\e}\bigg[\frac{\e(\veb_1-\veb_2)}{\zb_{12}}\mathcal{O}_{\Delta-2\e,+}(z_2,\zb)\nn\\
    &\qquad+(t\veb_1+(1-t)\veb_2)\partial_{\zb}\mathcal{O}_{\Delta-2\e,+}(z_2,\zb)-\veb~\partial_{\zb}\mathcal{O}_{\Delta-2\e,+}(z_2,\zb)\nn\\
    &\qquad+(\hb_1\partial_{\zb_1}\veb_1+\hb_2\partial_{\zb_2}\veb_2)-(\hb_1+\hb_2-\e)\partial_{\zb}\veb~\mathcal{O}_{\Delta-2\e,+}(z_2,\zb)\bigg]\bigg|_{\Delta = \Delta_1+\Delta_2-1}\\
    \circled{3} &= \frac{2\hat{c}_{\Gamma}}{z_{12}}\partial_{\Delta_1}\partial_{\Delta_2}\int_{0}^{1}dt~t^{\Delta_1-2}(1-t)^{\Delta_2-2}\bigg[\veb~\partial_{\zb}\mathcal{O}_{\Delta,+}(z_2,\zb)+(\hb_1+\hb_2)\partial_{\zb}\veb~\mathcal{O}_{\Delta,+}(z_2,\zb)\nn\\
      &\qquad~~-(\hb_1\partial_{\zb_1}\veb_1+\hb_2\partial_{\zb_2}\veb_2)\mathcal{O}_{\Delta,+}(z_2,\zb)-(t\veb_1+(1-t)\veb_2)\partial_{\zb}\mathcal{O}_{\Delta,+}(z_2,\zb)\bigg]\bigg|_{\Delta = \Delta_1+\Delta_2-1}~.
\end{align}
We first show that $\circled{1}$ vanishes. First note the identity $\partial_{\zb}\mathcal{O}(z_2,\zb) = \frac{1}{\zb_{12}}\partial_t\mathcal{O}(z_2,\zb)$, then by integrating the first and the last terms of \eqref{eq:term_1} by parts one can recast the integral in the following form
\begin{align}
    \circled{1} &= \frac{1}{z_{12}}\left(1-\frac{\pi^2}{6}\right)\int_{0}^{1}dt~t^{\Delta_1-2}(1-t)^{\Delta_2-2}\bigg[\partial_{\zb}\veb(\hb_1+\hb_2-1)-(\hb_1\partial_{\zb_1}\veb_1+\hb_2\partial_{\zb_2}\veb_2)\nn\\
    &~~+\frac{1}{\zb_{12}}\left(\frac{2\hb_2-1}{1-t}-\frac{2\hb_1-1}{t}\right)(\veb-t\veb_1-(1-t)\veb_2)+\frac{\veb_1-\veb_2}{\zb_{12}}\bigg]\mathcal{O}_{\Delta,+} \label{eq:term_1_modified}
\end{align}
we verify the vanishing of the above term for a variety of forms for $\veb_{1,2}$, they are summarised in the table below 
\begin{table}[ht!]
\begin{center}
    \begin{tabularx}{0.8\textwidth} { 
  | >{\raggedright\arraybackslash}X 
  | >{\centering\arraybackslash}X 
  | >{\raggedleft\arraybackslash}X | }
 \hline
 $\veb_1$ & $\veb_2$ & $\veb$ \\
 \hline
 1  & 1  & 1  \\
 \hline
 $\zb_1$   &  $\zb_2$ & $t\zb_1+(1-t)\zb_2$  \\
 \hline
 $\zb_1^2$ & $\zb_2^2$ & $(t\zb_1+(1-t)\zb_2)^2$\\
\hline
\end{tabularx}
\caption{}
\label{table:conformal_transformation_sl2r}
\end{center}
\end{table}
\\
for all the entries of table \ref{table:conformal_transformation_sl2r} it is easily seen that the integrand of \eqref{eq:term_1_modified} vanishes identically. Now similarly for term $\circled{2}$ we have 
\begin{align}
    \circled{2} &= \frac{\hat{c}_{\Gamma}}{z_{12}}\left(\frac{-\mu^2}{z_{12}\zb_{12}}\right)^{\e}\int_{0}^{1}dt~t^{\Delta_1-2-\e}(1-t)^{\Delta_2-2-\e}\bigg[-\partial_{\zb}\veb(\hb_1+\hb_2-\e-1)+(\hb_1\partial_{\zb_1}\veb_1+\hb_2\partial_{\zb_2}\veb_2)\nn\\
    &~~-\frac{1}{\zb_{12}}\left(\frac{2\hb_2-1-\e}{1-t}-\frac{2\hb_1-1-\e}{t}\right)(\veb-t\veb_1-(1-t)\veb_2)-(1+\e)\frac{\veb_1-\veb_2}{\zb_{12}}\bigg]\mathcal{O}_{\Delta-2\e,+} \label{eq:term_2_modified}
\end{align}
it also turns out that the above integrand vanishes for the transformations given in table \ref{table:conformal_transformation_sl2r}. Notice that the integrand of term $\circled{3}$ and $\circled{1}$ are identical and therefore by the virtue of this fact $\circled{3}$ vanishes identically. This concludes our proof of conformal invariance of \eqref{eq:loopOPEblock}. 

\
\bibliography{main}

\bibliographystyle{JHEP}

\end{document}